\documentstyle[12pt,aasms4]{article}

\def\power#1{\mbox{$\times10^{#1}\ $}}

\newcommand{\li}{$^{7}$Li }
\newcommand{\al}{$^{26}$Al }
\newcommand{\alg}{$^{26}$Al$^{g}$ }
\newcommand{\alm}{$^{26}$Al$^{m}$ }
\newcommand{\all}{$^{27}$Al }
\newcommand{\sod}{$^{22}$Na }
\newcommand{\sodd}{$^{23}$Na }
\newcommand{\neo}{$^{20}$Ne }
\newcommand{\mg}{$^{24}$Mg }
\newcommand{\mgg}{$^{25}$Mg }
\newcommand{\mggg}{$^{26}$Mg }
\newcommand{\msun}{M$_\odot$ }
\newcommand{\msyr}{M$_\odot$.yr$^{-1}$ }

\newcommand{\zaa}{A\&A~}
\newcommand{\zapj}{ApJ~}
\newcommand{\zanj}{AJ~}

\begin{document}

\title{New results on \al production in classical novae}

\author{Jordi Jos\'e}
\affil{Departament de F\'{\i}sica i Enginyeria Nuclear (UPC), Avinguda
V\'{\i}ctor Balaguer, s/n, E-08800 Vilanova i la Geltr\'u (Barcelona), SPAIN}
\author{Margarita Hernanz}
\affil{Institut d'Estudis Espacials de Catalunya (IEEC), 
CSIC Research Unit,
Edifici Nexus-104, C/Gran Capit\`a 2-4, 08034 Barcelona, SPAIN}
\and
\author{Alain Coc}
\affil{Centre de Spectrom\'etrie Nucl\'eaire et de Spectrom\'etrie de
Masse, IN2P3-CNRS, B\^at.104, F-91405 Orsay Campus, FRANCE}

\received{}
\accepted{}

\slugcomment{\underline{Submitted to}: \zapj Letters~~~~\underline{Version}:
\today}

\begin{abstract}

The production of \al by explosions of classical novae has been computed 
by means of a hydrodynamic code that follows both the accretion and the 
explosion stages. A special emphasis has been put on the analysis of the 
influence of the initial abundances of the accreted envelope, as well as 
on the nuclear reaction rates involved. With the most recent values of 
chemical composition and reaction rates available, \al production is lowered 
with respect to previous computations. According to our results, the final 
contribution of novae to the galactic \al is at most 0.4~\msun, which is a 
small part of the estimated \al in the Galaxy derived from COMPTEL 
observations of the 1809 keV emission. 

\end{abstract}

\keywords{novae, cataclysmic variables --- nuclear reactions, nucleosynthesis, 
abundances --- gamma-rays :theory}

\section{Introduction}

The discovery of \al in the interstellar medium by the HEAO-3 
satellite, through the detection of the 1809 keV $\gamma$-ray line 
(\cite{Mah82}, \cite{Mah84}), has raised the interest on potential 
$\gamma$-ray diagnostics of nucleosynthesis in several scenarios.
This $\gamma$-ray line is produced by the decay ($\tau\sim 10^6$ years) 
from the ground state of \al to the first excited state of \mggg, 
which de-excites to its ground state level by emitting a 1809 keV photon.
This detection has been 
confirmed by other space missions like the SMM $\gamma$-ray spectrometer     
(\cite{Sha85}) or several balloon-borne experiments.

Recent measurements made with the COMPTEL 
instrument on-board the Compton Gamma-Ray Observatory (CGRO) have provided 
a map of the 1809 keV emission in the Galaxy. This map shows an extended 
diffuse emission along the galactic plane, with a peculiar large scale 
asymmetry about the galactic center and a clumpy structure with several 
noticeable hot spots (\cite{Die95}).  The first interpretations of 
this map have suggested that novae and low--mass stars cannot be the major 
contributors to the 1809 keV emission, because their low individual yields 
and their high frequency should provide a smooth distribution, in 
contradiction with the irregular appearance of the emission (see \cite{PD96}
for a review). The general model suggested considers a two-component origin 
of the emission: a global component following the spiral pattern of the 
Galaxy (a presumed site of massive star formation) on which 
several localized regions of intense activity, such as the Vela region, 
are superimposed (\cite{CGD95}). However,
observations of four nearby supernova remnants with COMPTEL 
do not provide evidence for the 1809 keV emission, although the 
uncertainties in 
the distances do not allow to put severe constraints on \al production by 
supernovae (\cite{Kno96a}). Finally,
recent attempts to understand the 1809 keV map of the sky have analyzed in 
more detail its correlation with the spiral structure (\cite{Kno96b}). 
They derive a total \al mass of 2.5 \msun, from which at least 
0.7 \msun can be attributed to massive stars, stressing that they cannot 
exclude that a large fraction of \al is produced by novae or low--mass 
AGB stars. Therefore, with this still unclear panorama, it is worth 
studying the role played by classical novae in the synthesis of galactic \al. 

As pointed out by \cite{WF80}, \al production requires moderate peak 
temperatures, of the order of $\rm{T_{peak}} \leq ~2 \power{8}$ K,
and a fast decline from maximum temperature, conditions that are 
commonly achieved in nova outbursts. 
In the early 80s, one-zone model calculations of explosive hydrogen burning 
nucleosynthesis by \cite{HT82} and \cite{Wie86} suggested that classical 
novae might produce sufficient amounts of \al to account for some of the 
observed meteoritic anomalies but would not represent major galactic 
sources. 
But these calculations used only solar or CNO-enhanced envelopes.
New one-zone model calculations by \cite{WT90} and \cite{Nof91}, on the 
basis of ONeMg white dwarf stars, produced large amounts of long-lived 
radioactive nuclei, such as \sod and \al, concluding that these novae 
might represent important (though not dominant) sources of the galactic 
\al. However, according to \cite{Nof91}, no \al production results for models
with  T$_{\rm {peak}} \geq~2.7 \power{8}$ K.  Recent calculations 
have refuted this result (\cite{Pol95}, \cite{Sta93}), demonstrating the 
crucial role played by 
convection in order to carry some \al to the outer, cooler layers of the
envelope, where its destruction through proton captures is prevented.
But these fully hydrodynamical calculations involving ONeMg 
white dwarfs assume a composition for the white dwarf core (which is 
partially mixed with the envelope)
based on old calculations of C-burning nucleosynthesis by \cite{AT69} 
that need to be updated. Since the resulting nucleosynthesis is very 
sensitive to the envelope's composition, it is important to adopt a more 
realistic one. 
On the other hand, there are large uncertainties on some crucial reactions 
involving \al production. Thus, it is important to analyse the influence of 
the different prescriptions available on the final \al yields in the ejecta 
of classical novae. With this aim we have computed a series of hydrodynamic 
models of nova outbursts for several white dwarf masses and accretion rates, 
adopting updated initial chemical compositions and recent prescriptions for 
the crucial reaction rates. 
   
\section{Model and results}

A one-dimensional, lagrangian, implicit hydrodynamic code
has been developed, in order to follow both the hydrostatic accretion phase 
and the resulting hydrodynamic explosion (\cite{Jos96} and \cite{JH97}).
A time-dependent formalism for convective transport has been included
(\cite{W74}), where partial mixing in convective shells is treated by
means of a diffusion equation (\cite{PSS79}).
 With this code we have performed 
a series of computations of the complete evolution of accreting white 
dwarfs with masses ranging from 1 to 1.35 \msun, accreting at rates 
between 2\power{-8} and 2\power{-10}\msyr and with an initial luminosity of
10$^{-2}$ L$_{\odot}$ (see \cite{Her96} for an 
application of this code to the study of \li synthesis in novae). 
The matter transferred from the companion is assumed to be solar-like, but 
some process of mixing between the envelope and the outermost shells of 
the underlying core (shear mixing, diffusion) is 
assumed to take place. This assumption is based on the current prediction 
that enhanced CNO or ONeMg abundances are required in order to power the 
nova outburst and also to explain some of the observed abundances 
(\cite{LT94}. See also \cite{PK95} and \cite{Pol95} 
for recent calculations of CO and ONeMg novae, respectively). The problem 
of the initial composition of nova envelopes is complicated and far from 
being understood. A compromise solution is to adopt 
a mixture with 50\% by mass of core abundances.
This may be considered as representative of the degree of mixing, in view
of the mean metallicities observed in the ejecta of 'true' ONeMg novae
(\cite{LT94}. See \cite{KP96} for an analysis of the influence of the 
degree of mixing 
in the synthesis of \al). The composition of the underlying core has been taken 
from recent detailed evolutionary models, specially for the case of ONeMg white
dwarfs, which are the main contributors to \al production. These stars are 
made basically of $^{16}$O and \neo (\cite{Dom93} and \cite{Rit96}), 
since magnesium is almost absent (so we will call them ONe models). We 
want to stress that this issue has important consequences 
on the final yields of some species, like \al.

A complete reaction network has 
been directly coupled to the code to follow the detailed evolution of 
100 nuclear species, ranging from $^{1}$H to $^{40}$Ca, linked through 
more than 370 updated nuclear reactions. Both the ground and the isomeric 
states of \al have been included (\alg and \alm, respectively). 
The influence of nuclear uncertainties on \al yields 
has been investigated by \cite{Coc95}, with the use of the semi--analytical 
model of \cite{McD83}, for the cases considered by \cite{Pol95} (which have 
a high initial content of \mg, $\sim$0.1 by mass). 
They stressed the importance of the $E_p$=188~keV resonance 
in \alg(p,$\gamma)^{27}$Si which is not included in the 
\cite{CF88} rate (but whose strength was measured by \cite{Vog89}) and also 
of the large uncertainties still affecting the $^{25}$Al(p,$\gamma)^{26}$Si 
rate.
When \mg is the primary source of \al, formation of \alg through
$^{24}$Mg(p,$\gamma)^{25}${}Al($\beta^+)${}$^{25}$Mg(p,$\gamma)^{26}$Al$^{g}$
is bypassed by $^{25}$Al(p,$\gamma)${}$^{26}$Si($\beta^+)^{26}$Al$^{m}$,
at high temperature.  Another important reaction  for  $^{26}$Al
production is $^{27}$Al(p,$\alpha)^{24}$Mg which competes with
$^{27}$Al(p,$\gamma)^{28}$Si to partially recycle the initial Mg--Al material.
While the $^{27}$Al(p,$\gamma)^{28}$Si rate is sufficiently known in the
temperature range of interest, the $^{27}$Al(p,$\alpha)^{24}$Mg one is still
uncertain but appears to be lower than the one found in \cite{CF88}.
Since then, the resonances affecting this rate 
in the domain of temperature of novae have been studied by
\cite{Tim88} and \cite{Cha88}. Using two different techniques, they derive
similar upper limits for the strengths which are several orders of
magnitude lower than those used by \cite{CF88}.
Instead of being recycled in the Mg--Al region, the nuclear flow mainly
leaks out towards higher $Z$ elements, with the result of a reduced \al
production. Concerning the importance of the initial chemical composition, 
we stress that the sum X(\mg)+X(\mgg), and not 
only X(\mg), has to be considered as a seed for \alg production:
\mg is readily transformed into \mgg by the fast
\mg(p,$\gamma)^{25}$Al reaction followed by the beta decay of 
$^{25}$Al, provided that the leak through $^{26}$Si can be neglected.
Furthermore, unlike the CNO cycle, neither the NeNa nor the MgAl cycles are
expected to be closed and part of the initial neon and sodium will be
transferred to the MgAl region.
Indeed, in the temperature range $1.5 \leq \rm{T}_8 \leq 4$ the rate of the 
reaction $^{23}$Na(p,$\alpha)^{20}$Ne is approximately equal to that of
$^{23}$Na(p,$\gamma)^{24}$Mg according to \cite{CF88} or up to an order of
magnitude lower following a recent reevaluation by \cite{Eid95}.
Hence, a fraction of the initial X($^{23}$Na) will reach the MgAl
region providing additional seeds while the rest ends up as $^{20}$Ne. 
A fraction of $^{22}$Ne can also be transformed into $^{23}$Na.
At the temperatures prevailing in nova outbursts, the rate of
$^{20}$Ne(p,$\gamma$) is low and most of the initial $^{20}$Ne remains 
unburned.
However, in classical novae involving 
massive white dwarfs, a significant fraction of \neo can be transformed 
through the NeNa cycle and via $^{23}$Na(p,$\gamma)^{24}$Mg into $^{24}$Mg.
Accordingly, even in the absence of magnesium in the initial composition,
\al can be produced from $^{23}$Na, $^{22}$Ne or even some $^{20}$Ne in the 
most extreme cases. 

Snapshots of the evolution of several isotopes relevant to the \al
synthesis (\sodd, \mg, \mgg, \alg, \alm and \all)
are shown in figure 1, for a 1.25\msun ONe nova, with an accretion rate of 
2\power{-10}\msyr. The upper panel corresponds to the time
when the temperature at the burning shell has already reached $10^8$ K.
Nearly flat abundance profiles along the accreted envelope are found,
since the convective timescale is similar to the characteristic timescale
of most of the dominant nuclear reactions.
Proton capture reactions on \mg and \mgg provide the main source of the 
Al-group nuclei which increase their initial abundances. 
When the temperature in the burning shell rises up to 2.1\power{8} K 
(second panel), the nuclear timescale becomes much shorter than the 
convective
timescale. Hence, only partial mixing between adjacent convective shells
occurs and non-flat profiles are obtained. At this time, the abundance
of \al in the burning shell attains its maximum value, X(\al) 
$\sim$ 7.7\power{-3} by mass. \al is being progressively destroyed 
near the burning shell (panel 3) due to proton captures, since 
the seed nuclei \sodd, \mg and \mgg have almost been exhausted. 
The final abundances in the ejecta 
are shown in the lower panel, which corresponds to the time when the white
dwarf envelope has already expanded to a size of 
$\rm{R_{wd}} \sim 10^{12}$ cm.
The amount of \al in the envelope reflects the efficiency of convective 
transport, which 
carries \al isotopes to the outer, cooler layers of the envelope, where 
they can avoid destruction through (p,$\gamma$) reactions. 

The \al yields obtained in all models computed 
are summarized in table 1. All \al production appears in its ground state, 
since the isomeric state doesn't play any role. ONe novae
are more important producers than CO ones, due to the fact that
peak temperatures attained during CO nova outbursts are not high enough to
break the CNO cycle. Therefore, the production of \al comes from the
initial \sodd, \mg and \mgg.
It is worth noticing that the initial abundance of \mg is much lower than
the one adopted in the calculations performed by \cite{Pol95}, which were
based on the old nucleosynthesis
computations of carbon-burning by \cite{AT69}.
However, according to the above discussion, the presence of \mgg, \sodd and
$^{22}$Ne partially compensates this absence.

Only some combinations of peak temperatures around T$_{\rm{peak}} 
\sim$ 1--2\power{8} K and rapid evolution from maximum favor \al 
generation.  In this sense, explosion in low-mass ONe white
dwarfs, where the lower gravity provides softer explosions with lower
peak temperatures, favor  \al production. 
Our calculations (see table 1) show a minimum in the \al production near
$\rm {M_{wd}} = 1.25$ \msun, which fairly agrees with the trend  obtained 
in the semi-analytical analysis of \cite{Coc95}, but doesn't follow the 
monotonic decrease in the \al production as the white dwarf mass increases
reported by \cite{Pol95}. We point out, however, that the amount of \al
ejected into the interstellar medium decreases as the white dwarf mass or
the mass accretion rate increase, due to the fact that the mass of the 
envelope is smaller. It is also worth noticing that the amount of \al
present in the ejecta of the models described in this paper is systematically
lower than the mean values obtained by \cite{Pol95}. Two main
effects influence this result: first, a different initial composition (mainly
the lower initial content of Mg in our calculations), and second, the 
specific prescriptions adopted for the reaction rates of interest for the 
\al synthesis. 
The effect of the nuclear reaction rates has been
checked by means of 2 models of 1.25 \msun ONeMg white dwarfs, accreting
mass at a rate 2\power{-9} \msyr,
which are similar to one of the models
analyzed by Politano et al. (1995).
Both models differ in the specific
rates adopted for the reactions \al(p,$\gamma$)${}^{27}$Si and
${}^{27}$Al(p,$\alpha$)${}^{24}$Mg.
Since these reactions play no role during the accretion phase, the total
amount of mass accreted as well as the proper pressure at the onset of the
explosive phase are the same in both models. Hence, the differences 
obtained in the final yields reflect exactly the different prescriptions
for the reaction rates adopted.
When the rates from Vogelaar (1989) and Champagne et al. (1988) are
used, we obtain a mean value of X(\al) = 1.4 \power{-3} by mass
in the ejected shells, whereas X(\al) = 4.8 \power{-3} is found
when the Caughlan and Fowler rates (1988) are used 
(\cite{Pol95} obtained X(\al) = 9.45 \power{-3} by mass, for a similar
 model).
In addition, the ratio \al/\all increases from 0.4 to 3.0 in the second model.
This result is a direct consequence of the differences between the two 
prescriptions for the rate of ${}^{27}$Al(p,$\alpha$)${}^{24}$Mg:
the lower cross-section reported by Champagne et al. (1988) results in a less
efficient recycling of \all into \mg, which essentially will end up in 
the form of \al.

\section{Discussion and conclusions}

The production of \al by classical novae is very sensitive to the initial 
composition of the envelope and to the nuclear reaction rates adopted. 
ONe novae are more important \al producers than CO ones, because 
seed nuclei for the NeNa and MgAl cycles are almost absent in the latter ones. 
For the same reason, the amount of \al synthesized in ONe novae depends on 
the initial composition of the white dwarf core. When using the recently 
available chemical abundance profiles by \cite{Rit96}, a 
lower \al production is obtained. Furthermore, some improvements in the 
nuclear reaction rates since \cite{CF88} also lead to a lower \al production.  

The amount of \al injected into the interstellar medium by ONe nova events 
decreases as the mass of the underlying white dwarf increases. Hence, 
low-mass white dwarfs are  most likely candidates for \al production. 
Contribution from low-mass white dwarfs is favored both 
by the higher \al production and the higher ejected mass. But white dwarfs 
of masses lower than $\sim$1.1 \msun are expected to be CO white dwarfs, 
which are unable to produce important quantities of \al (see table 1).
Also, the observations of some Ne-novae, such as QU Vulpecula 1984, indicate 
high ejected masses ($\sim 10^{-3}$ \msun), unobtainable if a massive ONe 
nova is the responsible of the explosion. This has led some authors to 
propose a new scenario (\cite{Sha94}, \cite{SP94}):
low-mass CO white dwarfs, undergoing episodic accretion phases at high
rates, experience large metal enrichments from the ashes of He-burning. The 
subsequent nova explosion, correlated with a phase of a lower accretion rate, 
produces at the same time high \al abundances and high ejected masses.
But these models require a fine tuning of some parameters relative to the 
cataclysmic variable, such as the mass accretion rate.

A crude estimate of the total production of \al by novae can be computed 
by means of the expression from \cite{WT90}:

$$\rm M({}^{26}Al) = 0.4 { X({}^{26}Al) \over 2 \power{-3} } 
{frac(Ne) \over 0.25} {M_{ej} (M_\odot) \over 2 \power{-5} } 
{R_{nova} (events~yr^{-1}) \over 40}$$

\noindent
where X(\al) represents the mean mass fraction of \al in the ejecta, 
frac(Ne) the fraction of Ne-novae over the total number of classical
nova outbursts (between 0.25 and 0.53, but typically $\sim 1/3$, see 
\cite{TL86}, \cite{LT94}), $\rm M_{ej}$ the amount of 
mass ejected in an outburst, and $\rm R_{nova}$ the nova rate 
($\sim 46 \rm yr^{-1}$, \cite{HF87} or $\sim 20 \rm yr^{-1}$, \cite{DL94}).
Adopting these estimates of R$_{\rm{nova}}$ and frac(Ne) and 
our most favorable ONe nova case, 
for which $\rm M_{ej}$~(\al) = 1.7 \power{-8} \msun, we derive a
maximum contribution of classical nova outbursts 
to the galactic \al in the range 0.1 to 0.4 \msun. 
Although higher abundances of \al can be obtained for models with
higher enrichment from core material (i.e., 75\%), these may be
considered as very extreme cases, in view of the typical metallicities
observed in the ejecta of classical nova outbursts.
In summary, the contribution of novae to galactic \al
is small as compared with the one required to explain COMPTEL 
measurements (between 1 and 3 \msun). This is in agreement with the most 
accepted hypothesis of young progenitors as sources of galactic \al.

We want to stress that the two aspects that are crucial for the 
final \al yields are far from being understood; these are the process 
of mixing between the core and the envelope, which determines the initial 
chemical profile, and the exact rates of the 
nuclear reactions involved in \al synthesis. This latter topic will surely
be improved with the new compilation of nuclear reaction rates 
(\cite{NACRE}).

\acknowledgments
We are specially indebted to J. Isern, for valuable discussions about 
the topic of this paper. We acknowledge for partial support
the CICYT (ESP95-0091), the DGICYT (PB94-0827-C02-02), the CIRIT
(GRQ94-8001), the AIHF 95-335, the Human Capital and Mobility
Programme (CESCA/CEPBA) and the PICS 319.

\newpage
\begin{table}[htb]
\begin{center}
\caption{\al yields and ejected masses for some nova models}
\begin{tabular}{lllll}
Comp.                                &  M$_{\rm{wd}}$(\msun)     &
$\dot{{\rm M}}$(\msyr)               &  $\overline{\rm{X}}$(\al) &
M$_{\rm{ej}}$(\al)(\msun)\\
\tableline
CO  &  1.0   &  2\power{-10} & 1.6\power{-5}  & 3.5\power{-10} \\
CO  &  1.15  &  2\power{-10} & 4.7\power{-5}  & 6.2\power{-10} \\
ONe &  1.15  &  2\power{-10} & 9.3\power{-4}  & 1.7\power{-8 } \\
ONe &  1.25  &  2\power{-10} & 5.4\power{-4}  & 7.7\power{-9 } \\
ONe &  1.25  &  2\power{-8 } & 5.7\power{-4}  & 4.7\power{-9 } \\
ONe &  1.35  &  2\power{-10} & 7.2\power{-4}  & 3.1\power{-9 } \\
\hline
\end{tabular}
\end{center}
\end{table}

\newpage
\figcaption[fig1.eps]{
   Snapshots of the evolution of \sodd, \mg, \mgg,
   ${}^{26}$Al$^{g,m}$ and \all along the accreted envelope,
   for a 1.25 \msun ONe novae accreting at a rate 
   $\dot{\rm{M}}$ = 2 \power{-10} \msyr. 
   The successive panels, from top to bottom, correspond to
   the time for which the temperature at the burning shell reaches
   10$^{8}$, 2.1 \power{8}, $\rm{T_{max}} =$ 2.44 \power{8}, plus an 
   additional
   case for which the white dwarf envelope has already expanded to a size
   of $\rm{R_{wd}} \sim 10^{12}$ cm. The arrow indicates the base of the
   ejected shells.} 

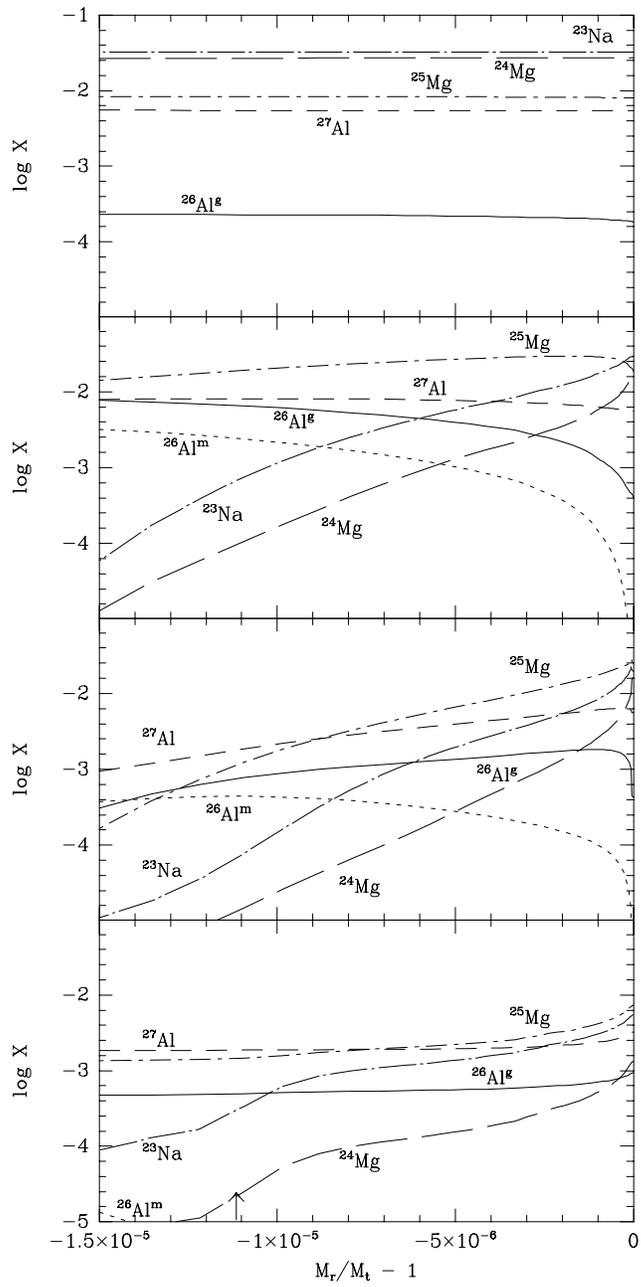
\end{document}